\documentclass[useAMS,usenatbib,usegraphicx]{mn2e}
\usepackage{times}
\usepackage{lscape}
\usepackage{amssymb}
\usepackage{color}
\def\cm3{$\rm cm^{-3}$}

\def\n0{$\rm n_{0}$}
\def\B0{$\rm B_{0}$}

\def\mc{$\mu$m}

\def\L12{L$_{12\mu m}$~}
\def\F12{F$_{12\mu m}$~}

\def\fe2{[Fe\,{\sc ii}]}
\def\s3{[S{\sc iii}]}
\def\h2{H$_{2}$}

\def\F{$F_{\lambda}$}

\title[Stellar Molecular Features in Nearby Galaxies]{The Stellar Spectral Features of Nearby Galaxies in the Near-Infrared: Tracers of Thermally-Pulsing Asymptotic Giant Branch Stars?}

\author[Riffel et al.]{Rog\'erio Riffel$^{1}$\thanks{E-mail:
riffel@ufrgs.br},
Rachel E. Mason$^2$,
Lucimara P. Martins $^6$,
Alberto Rodr\'{\i}guez-Ardila$^3$,
\newauthor
Luis C. Ho$^{4,5}$,
Rogemar A. Riffel $^7$,
Paulina Lira $^8$,
Omaira Gonzalez Martin $^{9,17}$,
\newauthor
Daniel Ruschel-Dutra $^{1,9}$,
Almudena Alonso-Herrero$^{10}$, 
Helene Flohic$^{11}$,
\newauthor
Richard M. McDermid$^{2,12,16}$, 
Cristina Ramos Almeida$^{9,17}$, 
Karun Thanjavur$^{13}$ and 
\newauthor
Claudia Winge$^{14}$
\\1) Departamento de Astronomia, Universidade Federal do Rio Grande do Sul. 
              Av. Bento Gon\c calves 9500, Porto Alegre, RS, Brazil.
\\2) Gemini Observatory, Northern Operations Center, 670 N. A'ohoku Place, Hilo, HI 96720, USA.
\\3) Laborat\'{o}rio Nacional de Astrof\'{i}sica - Rua dos Estados Unidos 154, Bairro das Na\c{c}\~{o}es.
CEP 37504-364, Itajub\'{a}, MG, Brazil.
\\4) Kavli Institute for Astronomy and Astrophysics,
Peking University, Beijing 100871, China.
\\5) Department of Astronomy, School of Physics, Peking University, Beijing 100871, China
\\6) NAT - Universidade Cruzeiro do Sul, Rua Galv\~ao
Bueno, 868, S\~ao Paulo, SP, Brazil.
 \\7) Universidade Federal de Santa Maria,
Departamento de F\'\i sica/CCNE, 97105-900, Santa Maria, RS, Brazil.
\\8) Departamento de Astronom\'ia, Universidad de Chile, Casilla 36-D, Santiago, Chile.
\\9)  Instituto de Astrof\' isica de Canarias, Calle V\' ia L\'actea, s/n, E-38205, La Laguna, Tenerife, Spain.
\\10) Instituto de Fisica de Cantabria, CSIC-UC, 39005 Santander, Spain.
\\11) University of the Pacific, Department of Physics, 3601 Pacific Avenue, Stockton, CA 95211, USA.
\\12) Australian Astronomical Observatory, PO Box 296, Epping, NSW 1710, Australia.
\\13) University of Victoria, Victoria, Canada.
\\14) Gemini Observatory, Southern Operations Center, c/o AURA, Casilla 603, La Serena, Chile.
\\16) Department of Physics and Astronomy, Macquarie University, NSW 2109, Australia.
\\17) Departamento de Astrof\' isica, Universidad de La Laguna, E-38205, La Laguna, Tenerife, Spain.}

\begin{document}

\date{}
\pagerange{\pageref{firstpage}--\pageref{lastpage}} \pubyear{}

\maketitle

\label{firstpage}

\begin{abstract}

We analyze the stellar absorption features in high signal-to-noise
ratio near-infrared (NIR) spectra of the nuclear region of 12 nearby
galaxies, mostly spirals. The features detected in some or all of the galaxies in this sample are
the TiO (0.843 \mc\ and 0.886 \mc),  VO (1.048 \mc), CN (1.1 \mc\ and 1.4 \mc), H$\rm _2$O (1.4 \mc\ and
1.9 \mc) and CO (1.6 \mc\ and 2.3 \mc) bands. The C$\rm _2$ (1.17
\mc\ and 1.76 \mc) bands are generally  weak or absent, although
C$\rm _2$ (1.76 \mc) may be weakly present in the mean galaxy
spectrum. A deep feature near 0.93 \mc, likely caused by CN, TiO and/or ZrO, is also detected in all objects. 
Fitting a combination of stellar spectra to the mean spectrum shows that the absorption features are produced by evolved stars: cool giants and supergiant stars in the early- or thermally-pulsing asymptotic giant branch (E-AGB or TP-AGB) phases. The high luminosity of TP-AGB stars, and the appearance of VO and ZrO features in the data, suggest that TP-AGB stars dominate these spectral features. 
However, a contribution from other evolved stars is also likely. Comparison with evolutionary population synthesis
models shows that models based on empirical libraries that predict relatively strong NIR features provide a more accurate description
of the data. However, none of the models tested accurately reproduces all of the features observed in the spectra.  To do so, the models will need to not only improve the treatment of TP-AGB stars, but also include good quality spectra of red giant and E-AGB stars. The uninterrupted wavelength coverage, high S/N, and quantity of features we present here will provide a benchmark for the next generation of models aiming to explain and predict the NIR properties of galaxies.

\end{abstract}


\section{Introduction} \label{intro}

Studying the unresolved stellar content of galaxies generally involves disentangling the various components contributing to the spectral energy distribution (SED), fitting a combination of simple stellar populations (SSPs) to derive information about age, metallicity, and star formation history. In the near-infrared (NIR, 0.85-2.5$\mu$m), the thermally pulsing asymptotic giant branch (TP-AGB) phase -- the last stage of the evolution of intermediate mass stars (M$\lesssim$6M$\odot$) -- is a particularly important component of the SSP models. These stars may be able to dominate the emission of stellar populations with ages $\sim$0.2 - 2 Gyr \citep[][see also \citet{mouhcine02} and \citet{dottori05}]{maraston05} being responsible for roughly half of the luminosity in the K-band \citep{salaris14}.

Unfortunately, a correct treatment of the TP-AGB phase is difficult to achieve, since the physics of the evolution of this stellar phase is still poorly known \citep{maraston05,marigo08,conroy10,conroy13,kriek10,zibetti13,noel13}. 
The complex processes occurring during this phase (mass-loss, changing opacities, dredge-up events, etc.) are difficult to accurately model, and must be anchored with empirical calibrations \citep[e.g.][]{marigo08}. The contribution of a population of TP-AGB stars to the light of a galaxy also depends on the duration of this evolutionary stage, which is not yet well constrained. In addition, the features of an observed spectrum of a TP-AGB star depend strongly on its age and metallicity \citep[e.g.][]{lancon01}, and we do not have NIR empirical spectra of stars covering the full range of these parameters expected in galaxies.  

These difficulties have led to the production of models making very different predictions for the NIR spectral region. In contrast to Evolutionary Population Synthesis (EPS) models based on isochrone synthesis, such as those of \citet[][hereafter BC03]{bc03}, the models of \citet{maraston05} and \citet[][hereafter M05 and M11]{maraston11}, based on the fuel consumption theory and using empirical spectra of C- and O-Rich stars for the TP-AGB phase, predict the presence of strong NIR molecular features, such as TiO (0.843 $\mu$m, 0.886 $\mu$m), CN (1.1 $\mu$m, 1.4 $\mu$m), C$\rm _2$ (1.17 $\mu$m, 1.76 $\mu$m), H$\rm _2$O (1.4 $\mu$m, 1.9 $\mu$m) and CO (1.6 $\mu$m, 2.3 $\mu$m) at the ages where these stars are abundant. When applied to galaxy spectra, the M05 models also predict much higher NIR luminosities than the BC03 ones \citep{maraston06}. The models are calibrated using Galactic and Magellanic Cloud globular clusters, and have been tested against a variety of astronomical data sets, with somewhat mixed results \citep{maraston06,riffel_clusters10,lyubenova10,kriek10,lyubenova12,melbourne12,zibetti13,marigo14}.

Given the challenges involved in constructing SSPs appropriate to modeling galaxy SEDs based solely on theoretical stellar spectra and/or limited empirical stellar spectra, searches for these bands in galaxies would be a valuable step forward in guiding the development and improvement of the models.  Some studies have been carried out in this area. The 1.1 $\mu$m CN band, for example, has been detected in AGN and starburst galaxies \citep{riffel07,ramos09}, along with the 1.4 $\mu$m CN band in a single object \citep{martins13}. The CO bandheads near 2.3 $\mu$m are well known, and the first detection of ZrO features at 0.8 - 1.0 $\mu$m was recently reported in a sample of starburst nuclei \citep{martins13}. Conversely, \citet{zibetti13} do not detect the TP-AGB spectral features predicted by M05 in their spectra of post-starburst galaxies at z$\sim$0.2.

Here we investigate the origin of the stellar molecular absorption features in NIR spectra of 12 nearby galaxies, and use the spectra to test two current EPS models. The detected bands include CN (1.1 $\mu$m, 1.4 $\mu$m), CO ($\sim$1.6 $\mu$m, $\sim$2.3 $\mu$m), H$_{2}$O ($\sim$1.4 $\mu$m, $\sim$1.9 $\mu$m), VO (1.048 $\mu$m) and TiO (0.843 $\mu$m, 0.886 $\mu$m). We also detect a feature at $\sim$0.93 $\mu$m which is probably due to a blend of CN, TiO, and ZrO bands. A tentative detection of a C$_2$ (1.76 $\mu$m) bandhead in the mean spectrum of the whole sample is also reported. While some of these bands have been discussed in previous work, the novelty of this study is the simultaneous detection and analysis of many features over a wide range of wavelengths.

\section{Sample and Data} \label{data}

\subsection{The Sample}
The galaxies in this paper (Table \ref{props}) are a subset of those observed
for the Palomar XD project \citep{Mason15}. That program
acquired high-quality nuclear NIR spectra of galaxies in the Palomar survey of nearby galaxies
\citep{ho95,ho97}, using the Gemini Near-Infrared Spectrograph
(GNIRS) on the Gemini North telescope \citep{elias06} . 
The GNIRS sample contains AGNs covering a wide range of luminosity and accretion rate. Of
these, a subsample was selected for this study on the basis of high
signal-to-noise ratio (S/N) data, strong stellar continuum, weak AGN
emission, and excellent cancellation of the strong telluric absorption coincident with the location of the CN and H$_2$O bands around 1.4 \mc. While some molecular bands are evident in the remaining 38 objects \citep{Mason15}, we restrict this discussion to the 12 galaxies with the highest-quality data. 

It is important to realize that these are not necessarily the galaxies where features due to TP-AGB stars would be expected to be the strongest. That is, no selection due to stellar population properties was made. The galaxies are simply those in which the data permit a close examination of the various bands that have been suggested to be signatures of TP-AGB stars.  As shown in Table~\ref{props}, with the exception of NGC~2832 and NGC~2768, all of the galaxies are spirals. Optical stellar population studies have been carried out for several of the galaxies, and in most cases old ages are suggested (Table~\ref{props}). Possible exceptions are NGC~2655, NGC~2768, and NGC~5005. The H$\beta$ line index in NGC2655 suggests a mean age of $\sim$2 Gyr \citep{sil06}, while studies disagree about whether old \citep{sil06,Zhang08} or intermediate-age \citep{Serra08,Crocker08} stars dominate in NGC~2768. In NGC~5005, \citet{cid04} find that an intermediate-age stellar population contributes about 45\% of the nuclear optical light.

Inter-comparing stellar ages obtained from different spectral regions, apertures, and line indices/stellar population models can be highly problematic. Nonetheless we use this information in \S\ref{discussion}, in which we draw some tentative conclusions about the type of stars causing the molecular absorption features observed in this galaxy sample.

\subsection{The data}

The spectra were obtained using GNIRS' cross-dispersed mode, which
provides simultaneous spectral coverage from $\sim$0.8-2.5 \mc. The
0.3\arcsec\ slit was used, providing R$\sim$1200, and was orientated
at the mean parallactic angle at the time of the observations. The
observations were executed in queue mode (Programs
GN-2012B-Q-80, GN-2013A-Q-16) and the observing condition criteria
allowed for observations to be taken with thin cirrus and seeing
$\lesssim$ 1\arcsec. The data were reduced using standard procedures, described in \citet{Mason15}.

Some stellar features of interest are located in regions of poor
atmospheric transmission, in which the telluric cancellation process
(dividing by a slightly shifted and scaled spectrum of an A star
observed near in time and airmass) may leave artefacts in the
spectrum. For instance, the 0.93 \mc\ feature lies close to
telluric H$\rm _2$O absorption at $\sim$0.95 \mc. The authenticity of this feature was verified in two ways. First, the shifting and
scaling applied to the standard star were adjusted, while verifying
that the strength and profile of the feature were not
significantly affected. Second, a standard star spectrum was reduced
using the same procedure applied to the galaxies, showing that no
similar feature was produced in the resulting spectrum (Fig. \ref{sources}). A similar procedure was applied to the H$\rm
_2$O absorption regions, indicating that the broad H$\rm _2$O bands
apparent in some objects are not artefacts of the data reduction
but true stellar features.

\begin{table*}
\centering
\caption{Galaxy Properties \label{props}}
\begin{tabular}{lclccccccc}
\hline
\hline
\noalign{\smallskip}
Galaxy      &Distance$^a$& Morphology$^a$ & AGN$^a$ & log $L(H\alpha)$$^a$  & Age (ref)$^b$ & Aperture$^c$ & SNR$^d$\\
                 & Mpc        &   &  &  erg s$^{-1}$    &  & pc &   \\
NGC 2655     & 24.4    & Sa             &  S2   &  39.55  & I (1) & 27$\times$82 &     118 \\    
NGC 2768     & 23.7    & S0/E6          &  L2   &  39.01  & I (1,2), O (3,4)& 27$\times$80 &     121 \\   
NGC 2832     & 91.6    & E3             &  L2   &  38.91  & O (2,5) & 135$\times$404 &   160  \\  
NGC 3147     & 40.9    & S(s)b I-II     &  S2   &  39.47  & O (6) & 54$\times$163 &    132 \\  
NGC 3718     & 17.0    & Sa             &  L1.9 &  38.46  & --  & 19$\times$58 &     123\\  
NGC 4548     & 16.8    & SB(sr)b I-II   &  L2   &  38.46  & O (7) & 9$\times$28  &     110 \\  
NGC 4565     &  9.7    & Sb             &  S1.9 &  37.97  & -- & 24$\times$73 &     100 \\  
NGC 4594     & 20.0    & Sab            &  L2   &  39.70  & O (8) & 21$\times$63 &     168 \\  
NGC 4750     & 26.1    & S(r)           &  L1.9 &  39.04  & -- & 31$\times$94 &     119 \\  
NGC 5005     & 21.3    & S(s)b II       &  L1.9 &  39.47  & I/O (9) & 19$\times$56 &     119 \\  
NGC 5371     & 37.8    & SB(sr)b I      &  L2   &  39.24  & -- & 50$\times$149 &    100 \\  
NGC 5850     & 28.5    & SB(rs)b I-II   &  L2   &  38.66  & O (10) & 50$\times$149 &    100 \\
\noalign{\smallskip}
\hline
\end{tabular}
\begin{list}{Table Notes:}
\item{(a) \citet{ho97} and references therein. (b) Stellar population age. O = old, age $\gtrsim$5 Gyr. I = intermediate-age, age $\sim$1-2 Gyr.
References: 1- \citet{sil06}. 2- \citet{Zhang08} 3- \citet{Serra08}. 4- \citet{Crocker08}. 5- \citet{Loubser09}. 6- \citet{bc03}. 7- \citet{sarzi05}. 8- \citet{sanchez06}.  9- \citet{cid04b}. 10- \citet {delorenzo13}}
(c) Slit width $\times$ extraction aperture of 0\arcsec.3 $\times$ 0\arcsec.9.
(d) Determined in the range 2.076-2.096 \mc.
    \end{list}
\end{table*}

%\clearpage

\section{Results}\label{results}

\subsection{The Spectral Features} \label{features}

\begin{figure*}
%\epsscale{1.0}
\includegraphics[scale=0.87,angle=90]{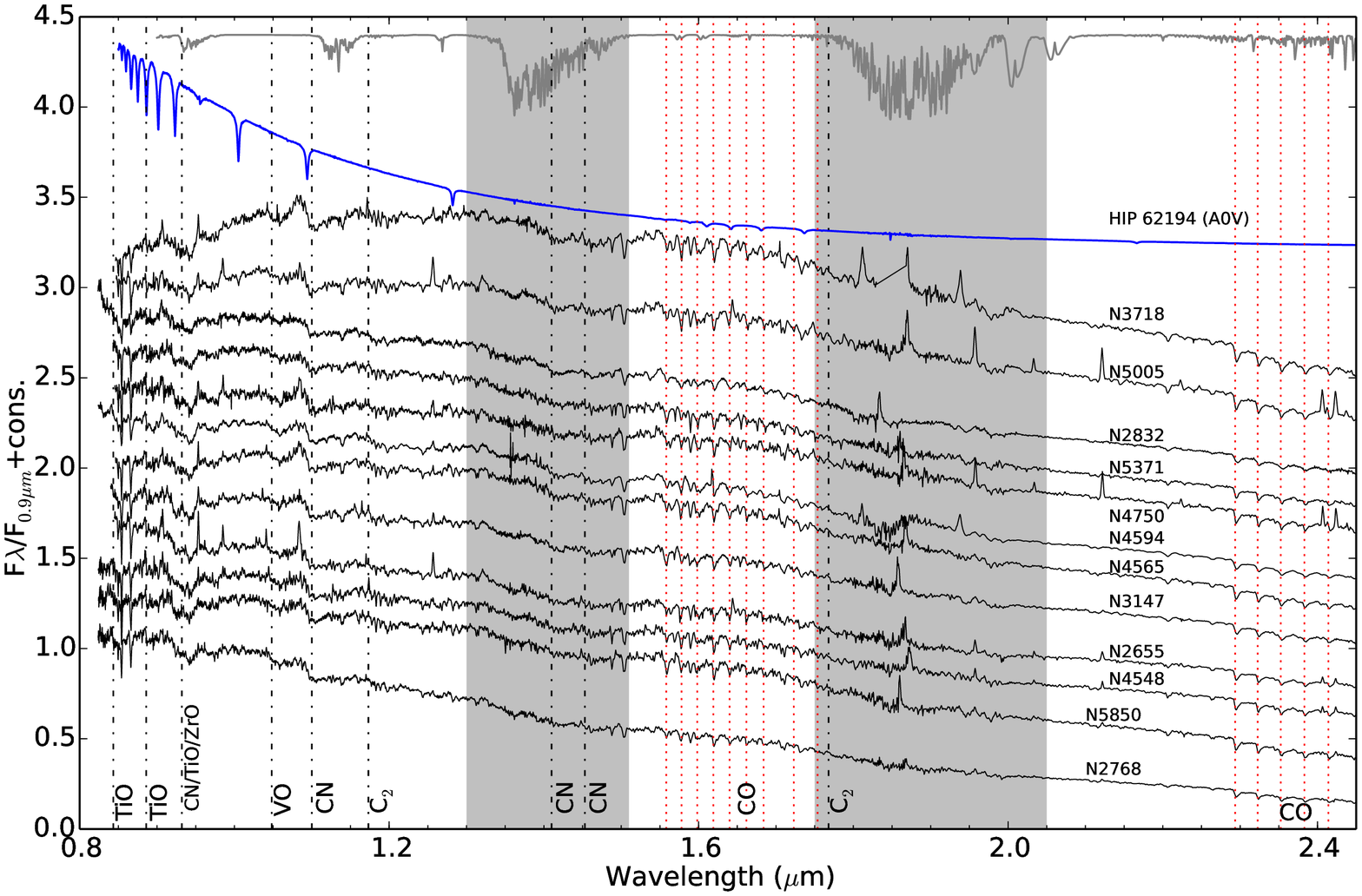}
%\plotone{figures/sources1.eps}
\caption[angle=0]{Rest frame spectra of the galaxy sample. Shaded regions indicate broad, stellar  H$\rm _2$O molecular bands, while dotted lines denote the CO bands. Other molecular bands are labeled at the bottom of the figure. A Mauna Kea atmospheric transmission spectrum for 1 mm H$_{2}$O and 1 airmass is also shown (gray), along with the spectrum of an A0 star (blue) reduced in the same manner as the galaxy data. Because of the small redshifts of the galaxies, the telluric features/residuals in the A0 star and atmospheric spectra do not correspond precisely to those in the galaxy data.}
\label{sources}
%\end{sidewaysfigure*}
\end{figure*}

\begin{figure*}
\includegraphics[scale=0.7]{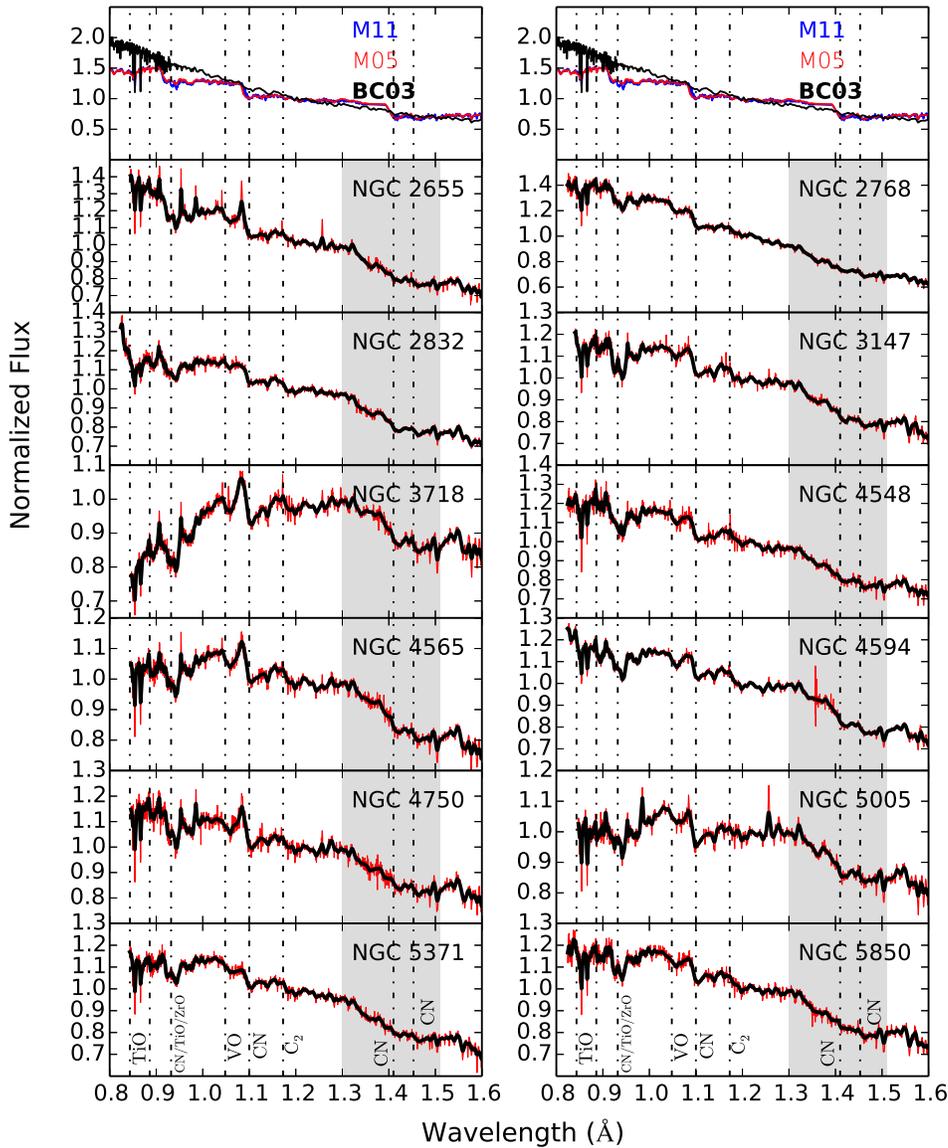}
\caption{Close-up of the 0.8 -- 1.6 $\mu$m region. Top: M11, M05 and BC03 models (\S\ref{ssp}). Other panels: observed spectrum (thin red line) and degraded to M05/BC03 spectral resolution (thick black line).}
\label{zoom1}
\end{figure*}

\begin{figure*}
\includegraphics[scale=0.7]{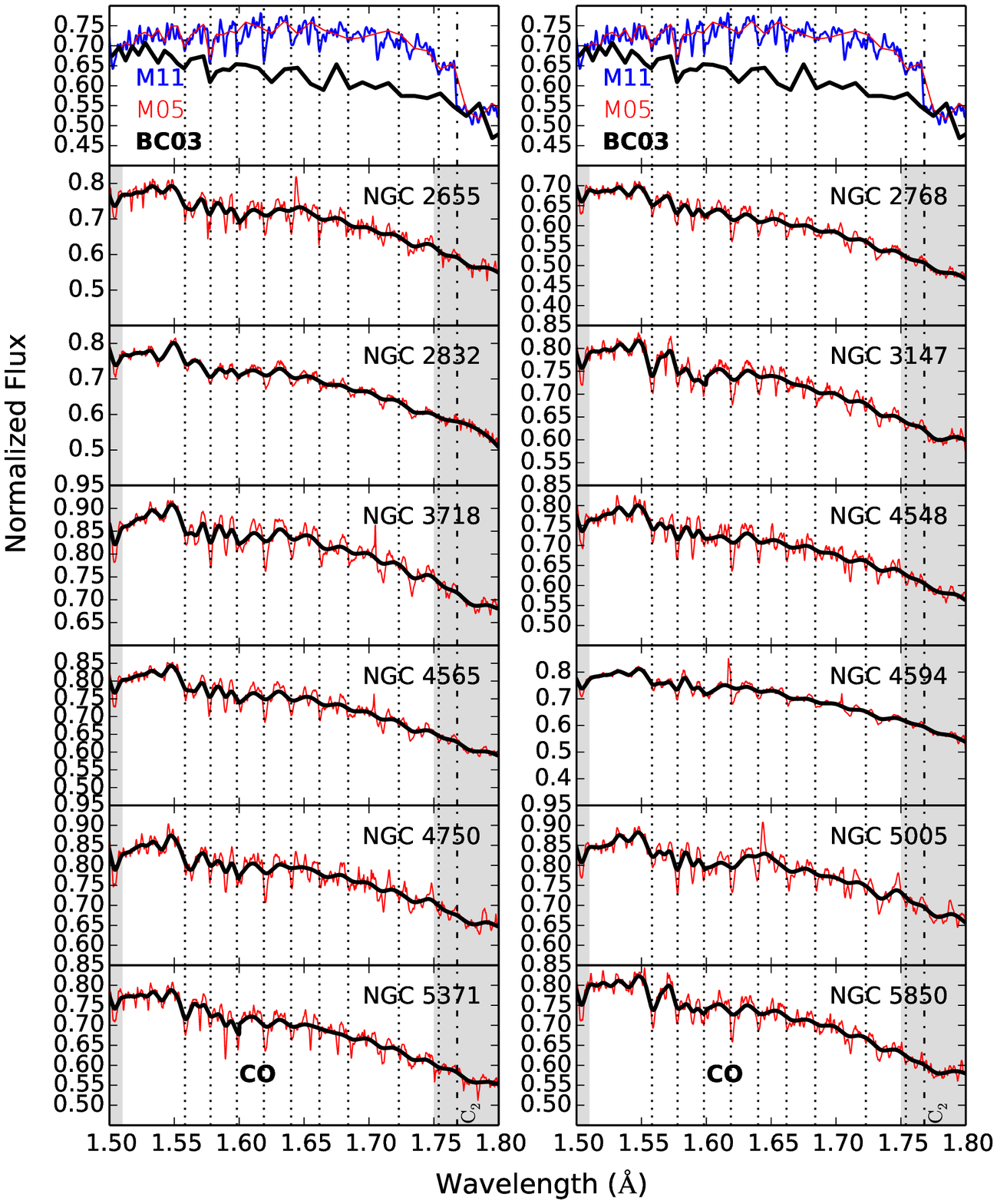}
\caption{Same as figure~\ref{zoom1}.}
\label{zoom2}
\end{figure*}

\begin{figure*}
\includegraphics[scale=0.7]{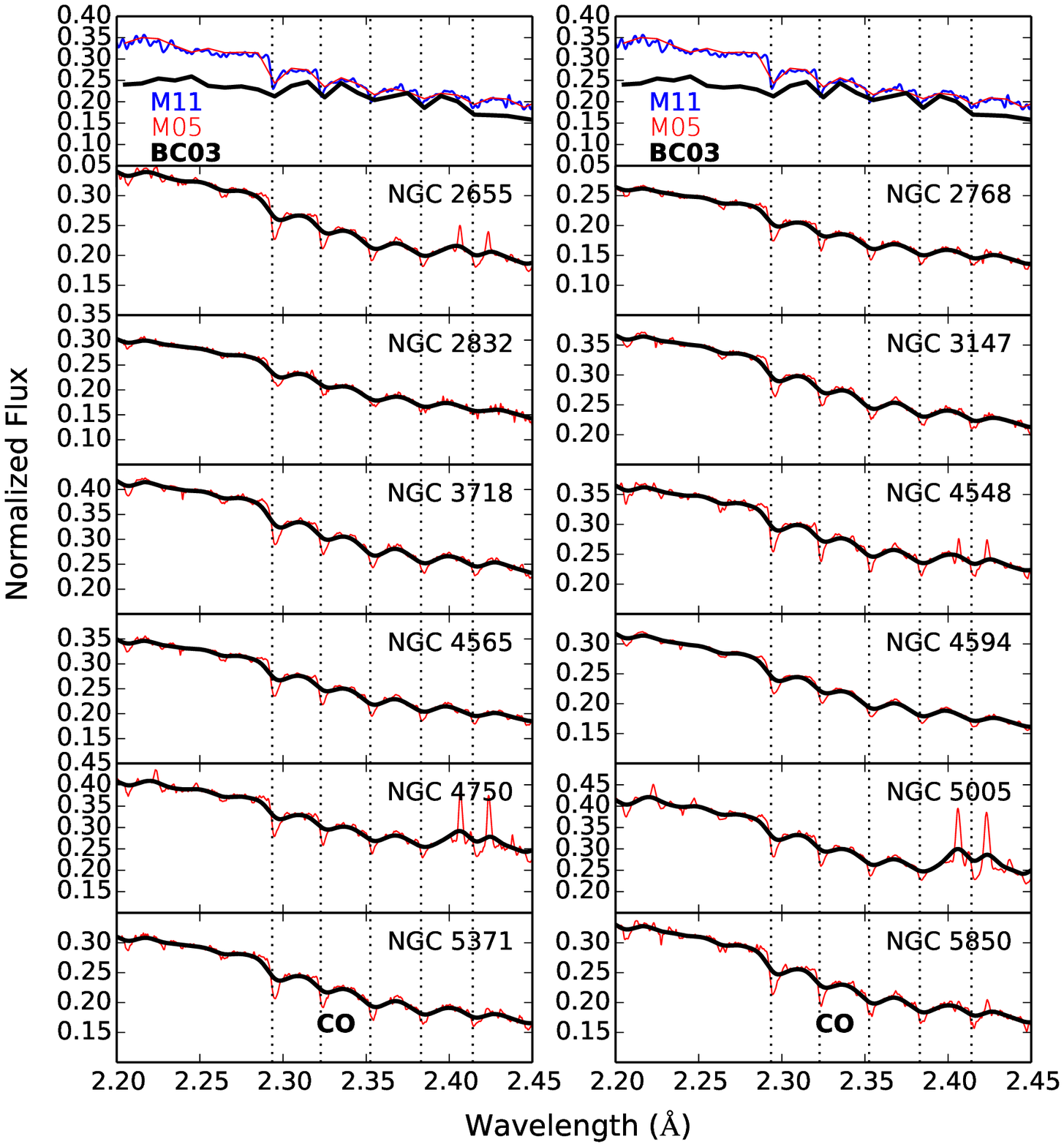}
\caption{Same as figure~\ref{zoom1}.}
\label{zoom3}
\end{figure*}

{The full set of spectra is shown in Figure \ref{sources}, while Figures \ref{zoom1} - \ref{zoom3} show close-ups of various regions of interest. As can be seen from these figures (see also Fig.~\ref{meanConts}), the} CN (1.1 \mc, 1.4 \mc) and CO (1.6 \mc, 2.3 \mc) bands are clearly detected in the majority of sources. While the 1.1 $\mu$m CN band is known in a number of galaxies \citep{riffel07}, the CN band at 1.4 \mc\ has to date only been detected in NGC~4102 \citep{martins13a}. The CN 1.4 \mc\ band falls in a region of poor atmospheric transmission and is therefore difficult to detect in low-redshift extragalactic sources. Thanks to the dry conditions under which these data were obtained, however, we detect the 1.4 \mc\ CN band in almost all the galaxies of the present sample.
According to M05, these bands are strong in TP-AGB stars, whose contribution to the integrated light peaks at intermediate ages.

The H$\rm _2$O 1.4 $\mu$m band is also detected in almost all sources, while the 1.8 \mc\  band is visible only in some (NGC~2655, NGC~2768, NGC~3147, NGC~4548, NGC~5850). C$_2$ lines are weak or absent in the individual galaxy spectra but the 1.76 \mc\  band may show up in the mean spectrum (see Fig.~\ref{hr}). These features are present in the IRTF atlas of cool stars  \citep[ see Fig.~7 to 34 of][see also \citet{cushing05}]{rayner09}, and are also predicted by M05 (their Figures 14 and 15). 

A comparison of the GNIRS spectra with the IRTF library shows that, besides the bands predicted by the M05 models, we also detect a VO feature at 1.048 \mc. A comparison of our Fig.~\ref{sources} with Fig.~8 of \citet{rayner09} shows that the VO band at 1.048 \mc, detected in many of the galaxies in our sample, most likely originate in late M~III stars in the TP-AGB phase. To our knowledge, this is the first time that these bands have been reported in galaxies. A broad feature ranging between $\sim$0.93 -- 0.95 \mc\ is also detected, but the identity of the carrier is uncertain. The feature may be due to a combination of CN (red system, $\Delta v = v''-v' = -1$), TiO ($\epsilon$ system, $\Delta v = -1$), and ZrO, which are all strong in the spectra of cool giant stars \citep{rayner09}. The 0.886 \mc\ band due to the $\delta$ system of TiO ($\Delta v = 0$) is detected in some of the spectra (NGC~2832, NGC~4594, and perhaps others). Depending on the spectral coverage, the 0.843 $\mu$m $\Delta v = 0$ bandheads of the $\epsilon$ system of TiO are also detected in those objects, blended with the well-known Calcium triplet lines. 

We have measured the equivalent widths (EW) of the strongest molecular bands, including the 0.93 $\mu$m CN/TiO/ZrO and 1.048 $\mu$m VO bands that are particularly relevant to TP-AGB stars. The measurements were performed using an updated version of the {\sc pacce} code \citep{riffel_vale11}. In this code, the continuum uncertainty is taken as the root mean squere of the difference between the linear fit to the pseudo continuum and the observed one. Then a Gaussian distribution of errors, using the above difference as 1-sigma standard deviation, is assumed to simulate new pseudo-continua. Finally, EW uncertainties are assumed to be the standard deviation of 500 measurements of EW using the simulated pseudo continua as continuum points for the linear fit. The line definitions used are listed in Table~\ref{ewdefs} and the measured values are in Table~\ref{ew}.

%\begin{sidewaysfigure*}

%\subsection{Mean spectrum }\label{meanspec}

As the spectra of the 12 galaxies have rather similar continuum and emission line properties (see Fig.~\ref{sources}), we averaged them to construct a spectrum with very high S/N, allowing further verification that the detected features are real. The spectra were normalized to unity at 1.223 \mc\ before computing the mean value of the fluxes, pixel-by-pixel. The resulting spectrum (Fig.~\ref{hr}) supports the conclusion that the features observed in the individual galaxies are real, since they clearly appear in the mean spectrum. As  the galaxies are at different redshifts, when they are shifted to the rest frame to create the mean spectrum, any systematic effects are shifted to different positions and will therefore tend to cancel out or be weakened in the mean spectrum. Additionally, (i) reducing a standard star using the same procedures results in a very clean spectrum with few weak telluric line residuals (\S\ref{data}), and (ii) we can fit the observed mean absorption spectrum very well using the IRTF stellar library (\S\ref{discussion}), thus indicating that the features are real and due to stars.

\begin{table}
\centering
\caption{Line limits and continuum bandpasses. \label{ewdefs}}
\begin{scriptsize}
\begin{tabular}{lcc}
\hline
\hline
\noalign{\smallskip}
ID            & line limits   & continuum bandpass \\
              & (\mc)         &  (\mc) \\
\hline
\noalign{\smallskip}            
CN/TiO/ZrO     & 0.9200-0.9500   & 0.8760-0.8800,0.9570-0.9650\\
VO             & 1.0470-1.0650   & 1.0350-10390, 1.0660-1.0740\\
CN11 ($\alpha$)& 1.0900-1.1300   & 1.0660-10740,1.1310-1.1360\\
CN14a          & 1.4050-1.4300   & 1.3950-13990, 1.4340-1.4390, 1.4750-1.4840\\
CN14b          & 1.4500-1.4715   & 1.3950-13990, 1.4340-1.4390, 1.4750-1.4840\\
CO16a          & 1.5735-1.5810   & 1.5110-15170, 1.5390-1.5410, 1.6270-1.6310, 1.6570-1.6580\\
CO16b($\alpha$)& 1.6175-1.6285   & 1.5110-15170, 1.5390-1.5410, 1.6270-1.6310, 1.6570-1.6580\\
CO22a($\alpha$)& 2.2910-2.3200   & 2.2690-22790, 2.3950-2.3999\\
CO22b($\alpha$)& 2.3200-2.3420   & 2.2690-22790, 2.3950-2.3999\\
CO22c($\alpha$)& 2.3420-2.3695   & 2.2690-22790, 2.3950-2.3999\\
\hline
\end{tabular}
\end{scriptsize}
\begin{list}{Table Notes:}
\item $\alpha$ based on \citet{riffel09} and \citet{riffel_clusters10}.
\end{list}
\end{table}

\begin{table*}
\renewcommand{\tabcolsep}{1.1mm}
\caption{Equivalent widths of the absorption bands in the galaxy sample. \label{ew}}
\begin{tabular}{lcccccccccc}
\hline
\hline
\noalign{\smallskip}
Galaxy    &   CN/TiO/ZrO      &   VO           &   CN11           &   CN14a          &   CN14b   &   CO16a   &   CO16b   &   CO23a   &   CO23b   &   CO23c\\
\hline
\noalign{\smallskip}            
NGC~2655  &  20.20$\pm$0.50  &  3.66$\pm$0.37  &  12.06$\pm$0.52  &  8.31$\pm$0.50  &  4.71$\pm$0.34  &  4.12$\pm$0.13  &  6.84$\pm$0.19  &  23.31$\pm$0.24  &  22.60$\pm$0.30  &  29.83$\pm$0.42\\
NGC~2768  &  17.23$\pm$0.53  &  3.75$\pm$0.24  &  11.14$\pm$0.46  &  5.09$\pm$0.58  &  5.88$\pm$0.37  &  3.23$\pm$0.07  &  6.45$\pm$0.11  &  23.97$\pm$0.27  &  22.61$\pm$0.22  &  30.42$\pm$0.33\\
NGC~2832  &  16.16$\pm$0.49  &  2.17$\pm$0.35  &  11.24$\pm$0.51  &  8.88$\pm$0.34  &  5.00$\pm$0.27  &  4.30$\pm$0.10  &  4.85$\pm$0.27  &  21.59$\pm$0.43  &  21.24$\pm$0.43  &  30.63$\pm$0.68\\
NGC~3147  &  23.80$\pm$0.41  &  3.23$\pm$0.46  &  9.45$\pm$0.51   &  6.68$\pm$0.42  &  3.62$\pm$0.38  &  2.39$\pm$0.08  &  5.87$\pm$0.16  &  18.92$\pm$0.21  &  17.44$\pm$0.29  &  24.18$\pm$0.40\\
NGC~3718  &  18.34$\pm$0.58  &  4.92$\pm$0.32  &  9.07$\pm$0.61   &  8.22$\pm$0.49  &  4.92$\pm$0.39  &  3.37$\pm$0.09  &  6.28$\pm$0.19  &  22.36$\pm$0.33  &  20.58$\pm$0.26  &  27.90$\pm$0.37\\
NGC~4548  &  22.58$\pm$0.75  &  3.94$\pm$0.45  &  10.49$\pm$1.15  &  6.33$\pm$0.70  &  4.74$\pm$0.47  &  3.33$\pm$0.10  &  6.90$\pm$0.17  &  20.33$\pm$0.25  &  18.77$\pm$0.25  &  25.97$\pm$0.36\\
NGC~4565  &  16.61$\pm$0.44  &  3.43$\pm$0.24  &  9.01$\pm$0.49   &  7.63$\pm$0.49  &  4.16$\pm$0.31  &  3.64$\pm$0.08  &  6.66$\pm$0.17  &  21.98$\pm$0.25  &  20.94$\pm$0.24  &  28.45$\pm$0.42\\
NGC~4594  &  20.07$\pm$0.40  &  4.81$\pm$0.24  &  8.07$\pm$0.40   &  10.43$\pm$1.02 &  3.74$\pm$0.37  &  3.41$\pm$0.07  &  3.68$\pm$0.16  &  23.80$\pm$0.16  &  22.13$\pm$0.16  &  30.86$\pm$0.32\\
NGC~4750  &  16.44$\pm$0.61  &  3.98$\pm$0.40  &  8.32$\pm$0.56   &  5.37$\pm$0.55  &  4.43$\pm$0.33  &  1.96$\pm$0.08  &  6.41$\pm$0.15  &  20.86$\pm$0.28  &  20.61$\pm$0.30  &  26.45$\pm$0.48\\
NGC~5005  &  11.86$\pm$0.41  &  1.70$\pm$0.32  &  10.52$\pm$0.53  &  7.93$\pm$0.23  &  3.39$\pm$0.27  &  4.28$\pm$0.08  &  6.15$\pm$0.13  &  22.80$\pm$0.22  &  23.16$\pm$0.26  &  30.06$\pm$0.48\\
NGC~5371  &  16.80$\pm$0.48  &  1.75$\pm$0.34  &  7.26$\pm$0.67   &  4.84$\pm$0.38  &  3.81$\pm$0.35  &  1.94$\pm$0.12  &  6.41$\pm$0.21  &  22.65$\pm$0.25  &  20.62$\pm$0.28  &  28.82$\pm$0.44\\
NGC~5850  &  18.45$\pm$0.51  &  3.11$\pm$0.39  &  10.06$\pm$0.66  &  3.85$\pm$0.44  &  3.87$\pm$0.31  &  3.07$\pm$0.09  &  6.72$\pm$0.14  &  22.58$\pm$0.30  &  21.94$\pm$0.26  &  29.15$\pm$0.34\\
\hline
\end{tabular}
\end{table*}

\subsection{Stellar population synthesis} \label{ssp}

Bands from carbon-containing molecules are present in the galaxy spectra in Fig. \ref{sources}, along with VO and possibly a contribution from ZrO in the band at 0.93 $\mu$m. The presence of these molecules suggests a significant contribution from TP-AGB stars to the NIR spectra of the galaxies. To further investigate (1) whether these molecular bands really are due to TP-AGB stars, and (2) whether the available stellar population synthesis models are able to reproduce them, we carry out a stellar population synthesis on the mean spectrum using the {\sc starlight} spectral synthesis code
\citep{cid04,cid05a,cid05b,asari07}. The procedures followed are described in \citet{riffel09} and \citet{dametto14}, including, for example, the methods used to handle extinction, emission lines and differences in spectral resolution. Briefly, {\sc starlight} fits an observed spectum $O_{\lambda}$ with a combination, in different proportions, of
$N_{*}$ SSPs, solving the equation:
\begin{equation}
M_{\lambda}=M_{\lambda 0}\left[\sum_{j=1}^{N_{*}}x_j\,b_{j,\lambda}\,r_{\lambda} \right] \otimes G(v_{*},\sigma_{*}),
\end{equation}
where  $M_{\lambda}$ is a model spectrum, $b_{j,\lambda}\,r_{\lambda}$ is the reddened spectrum of the $j$th SSP normalised at
$\lambda_0$; $r_{\lambda}=10^{-0.4(A_{\lambda}-A_{\lambda 0})}$ is the reddening term; $M_{\lambda 0}$ is the
theoretical flux at the normalisation wavelength; $\vec{x}$ is the population vector, and $G(v_{*},\sigma_{*})$ is the Gaussian distribution used to model the line-of-sight stellar
motions, which is centred at velocity $v_{*}$  with dispersion  $\sigma_{*}$.
The final fit is carried out searching for the minimum of the equation:

\begin{equation}
\chi^2 = \sum_{\lambda}[(O_{\lambda}-M_{\lambda})w_{\lambda}]^2,
\label{equchi}
\end{equation}
where emission lines and spurious features are masked out by fixing $w_{\lambda}$=0. The quality of the fit is accessed by $\chi^2_{Red}$ wich is the $\chi^2$ given by equation~\ref{equchi} divided by the number of points used in the fit and by $adev = |O_{\lambda} - M_{\lambda}|/O_{\lambda}$, which is the percentage mean deviation over all fitted pixels. For a detailed description of {\sc starlight} see its manual\footnote{http://astro.ufsc.br/starlight/}. For applications in the optical see \citet{cid05a}, and in the NIR see \citet{riffel09}.

We use 4 different approaches to define the base set of spectra used by the code, as follows: 
i) {\it stars approach}: a base composed of all the 210 dereddened stars (spectral types F--S/C, F being the hottest available) in the IRTF spectral library
\citep{rayner09,cushing05} and 
ii) {\it stars no TP-AGB approach}: the same as in (i) but removing all the TP-AGB stars from the base.
iii)  {\it M11 approach}: EPS models of M11 (an example of models including empirical spectra of C- and O-Rich TP-AGB stars),
which consist of theoretical SSPs covering 12 ages ($t=$\,0.01,0.03, 0.05, 0.1, 0.3, 0.5, 0.7, 1, 2, 5, 9 and 13\,Gyr) at solar
metallicity; 
iv) {\it BC03 approach}: SSPs from BC03 (an example of models calibrated with broad band photometry) with the same ages and metallicity as 
above.

\section{Discussion}\label{discussion}

\subsection{Origin of the features}\label{origin}

To link the molecular absorption features to the kind of star in which they may arise, we first revisit some relevant aspects of the later stages of stellar evolution. 

Once the helium in the core of a star is exhausted, He burning begins in a shell surrounding the core. This marks the arrival of the star on the asymptotic giant branch (AGB). 
The star starts to expand, becoming cool and luminous, and develops a growing convective outer region. At this stage the second dredge-up occurs, mixing the end points of helium burning, mainly C and O, into the star's atmosphere. At the end of this phase the s process becomes the dominant mechanism of nuclear fusion, producing elements such as Zr and V \citep{habing04}.

The star then starts to undergo thermal pulses, leaving the E-AGB stage and entering the TP-AGB phase. The third dredge-up takes place, and processed material from the star's interior is transported into its atmosphere.  At this point the spectrum of the star will present evidence of enhanced C and O abundance, along with features due to species containing the newly-formed Zr and V \citep{habing04}. For example, the NIR CN absorption band is particularly enhanced in carbon stars, where there is residual carbon that is not bound with oxygen in CO molecules. As N type carbon stars are produced after the third convective dredge-up along the TP-AGB phase \citep{habing04}, the detection of deep CN features may indicate stellar populations rich in this kind of star \citep{lancon01,riffel07}. Similarly, VO and ZrO are produced with the residual oxygen not used in the formation of CO and TiO. Since V and Zr are only produced by the s process, these features are normally only detected in TP-AGB stars. 

%The expectations of stellar evolutionary theory are borne out by the observations. 
Below, we relate the spectral features in the galaxies to those observed in various kinds of star. It is therefore necessary to understand the stellar characteristics (spectral types and other traits) that identify a star as being in the TP-AGB phase. Most AGB stars have spectral classes M, S, and C \citep{habing04}. Of the two main types of carbon star, C-N stars are in the TP-AGB phase while C-R stars are not. All the molecular bands in C stars are from carbon-containing molecules. The spectra of early M-type stars contain TiO bands, while VO becomes apparent at later spectral types. S stars, however, are characterized by distinctive ZrO absorption. As noted above, s-process elements such as Zr and V are normally only present in AGB stars in the thermally-pulsing stage. 

TP-AGB stars can also be distinguished by the variability caused by their long-period pulsations. The stellar library of \citet{rayner09} includes several IR spectra of TP-AGB stars, identified by their characteristic variability types (irregular, semi-regular, and Mira), and their data illustrate the above trends. Inspecting their series of dwarf, giant and supergiant spectra (their Figures 7, 8, and 9) shows that the only molecular features detected in the dwarf stars are TiO, CO, H$_2$O and FeH. CN bands are present in the giants and supergiants, but are most prominent in the late-type supergiants with TP-AGB-like variability characteristics (the K1 Ia-Iab, K5 Ib, M3 Iab-Ia, and M5 Ib-II stars in their Figure 9). ZrO also becomes prominent in the coolest of the TP-AGB supergiants in that figure. In the latest-type of the TP-AGB giants (M4 III, M6 III, M7 III, and M8-9 III in Figure 8 of Rayner et al. 2009), absorption from the VO molecule becomes apparent.

\begin{figure*}
%\epsscale{1.2}
\includegraphics[scale=0.7]{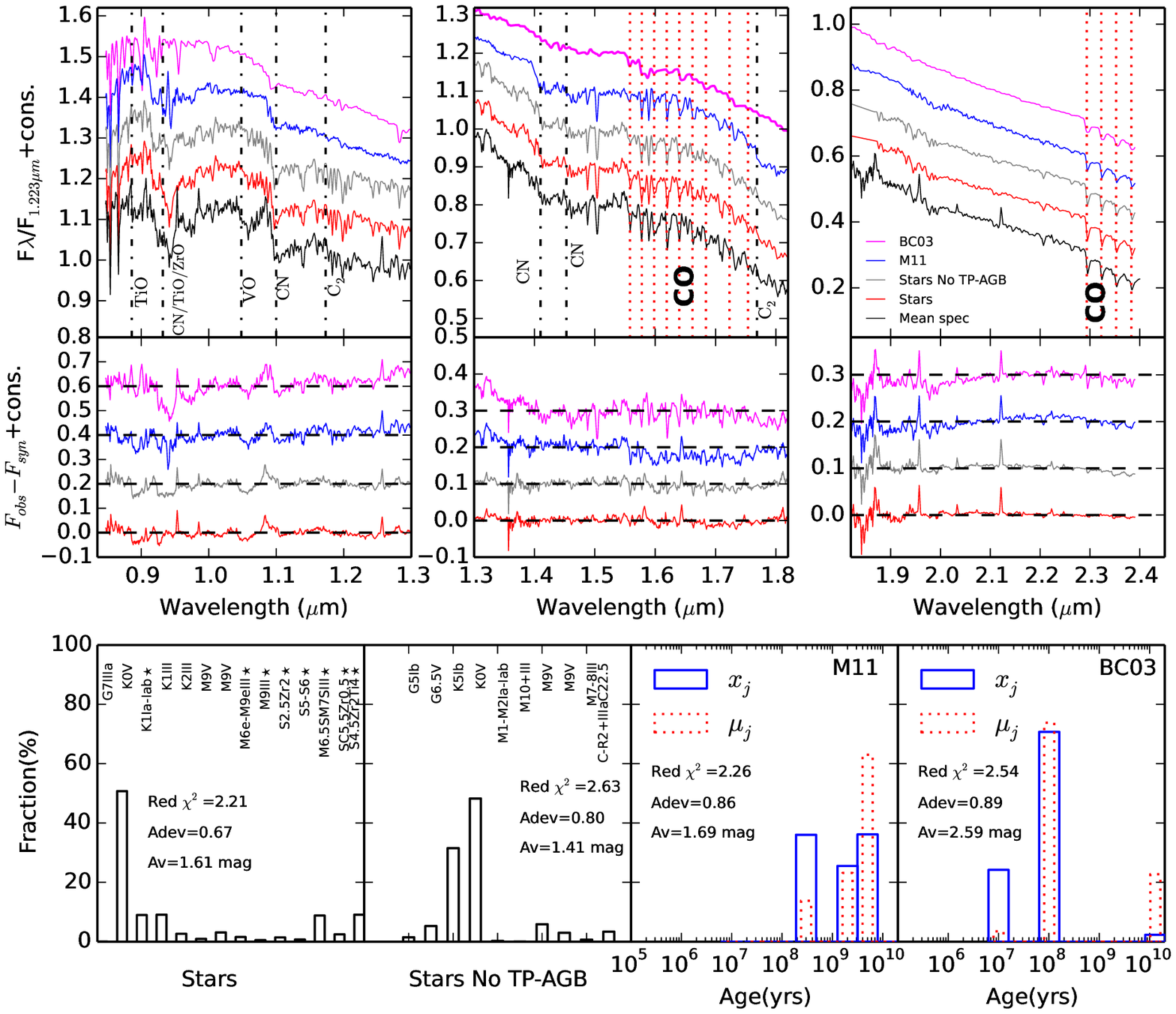}
\caption{Top: Mean spectrum (black), along with EPS fits using  (lowest to highest) the IRTF stars, stars no TP-AGB, M11 and BC03 models. Middle: the difference between the mean spectrum and its fits, following the same order as in the top. Bottom: synthesis results from these three methods. $x_j$ denotes light fractions and $\mu_j$ denotes mass fractions. Stars spectral types are taken from \citet{cushing05,rayner09} and $\star$ denotes TP-AGB stars. The goodness of the fits and $Av$ are on the labels. For display purposes we removed from the histograms stars in the base set that do not contribute to the fits.}
\label{hr}
\end{figure*}

\begin{figure*}
\includegraphics[scale=0.7]{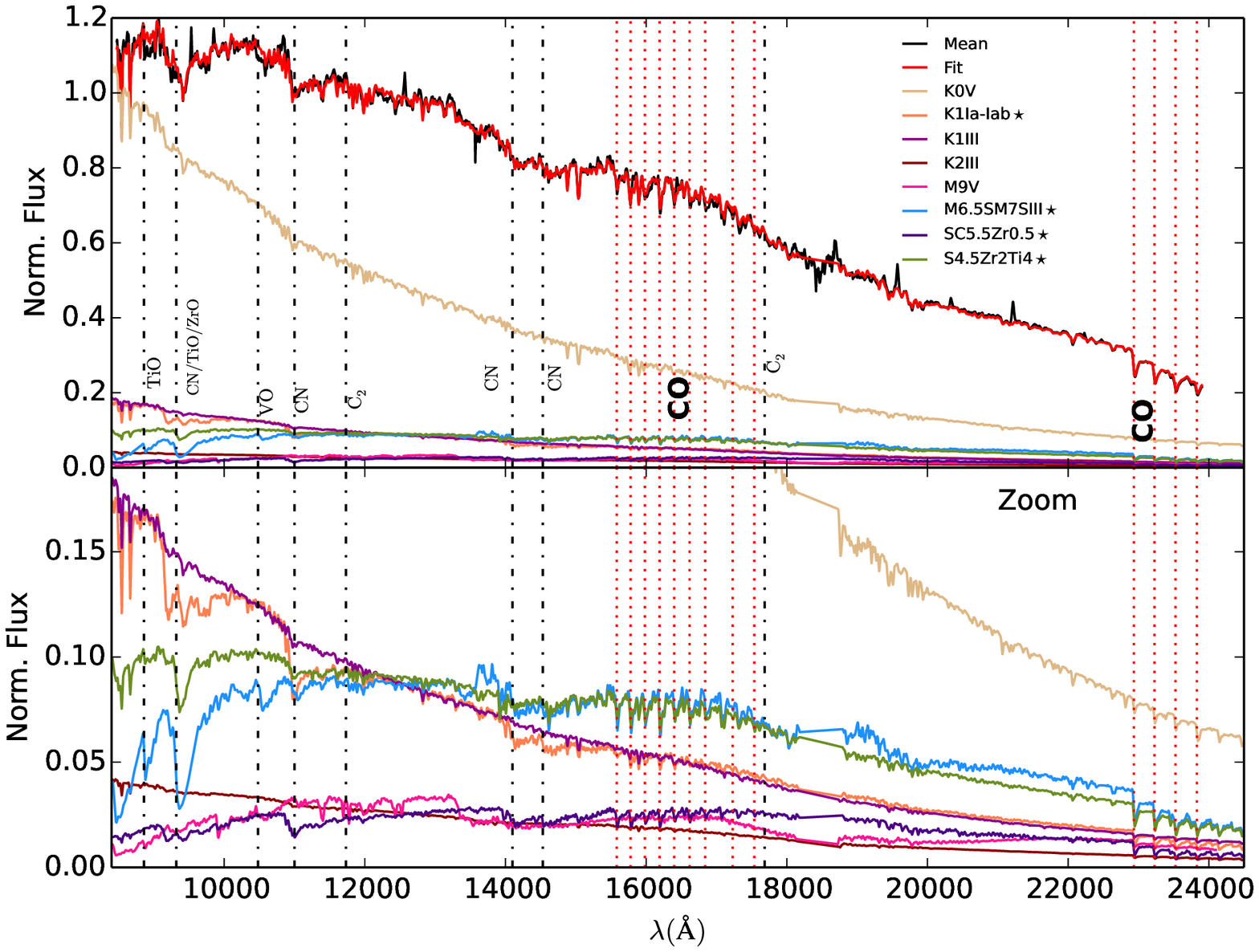}
\caption{Fractional contribution of the stars used in the fit using the {\it stars approach} (Top). For display purposes we removed the stars contributing with less than 2\% (to see the contribution of stars contributing with less than 2\% see Fig.~\ref{hr}) from the plot. The contribution of the TP-AGB stars (marked with $\star$) to the observed absorption features is clearly observed, a zoom to these features is shown at the bottom. For more details see Sec~\ref{ssp}.}
\label{meanConts}
\end{figure*}

The results of the stellar population synthesis are shown in Figure \ref{hr}. When using the {\it stars approach} the spectrum can be well reproduced, and a significant fraction ($\sim$ 35\%) of the light comes from TP-AGB stars (the stars marked with $\star$ in the lower left panel of the figure).{\bf This is illustrated by Figure~\ref{meanConts}, which shows that these stars account for the molecular features particularly at the short wavelength end of the spectrum.} However, when the TP-AGB stars are removed, the code replaces them with a variety of E-AGB and red giant stars, notably a large contribution from a K5Ib star and a significant contribution from a C-R type carbon star. The quality of the fit is similar overall, with just a small increase in the $\chi^2$ and $Adev$\footnote{The percentage mean $\rm | O_{\lambda} - M_{\lambda} | /O_{\lambda}$ deviation over all fitted pixels.} values. 

Taken at face value, this would suggest a degeneracy between R-type carbon rich stars summed with E-AGB stars, and the TP-AGB stars. However, there are two important details that have to be taken into account. First, we are trying to fit a galaxy's stellar population with the observed spectra of stars from a library which is neither complete nor absolutely flux-calibrated, and most importantly, not constrained by the rules of stellar evolution, which would require the use of an isochrone and an assumption about the initial mass function.This would be the correct approach to create an SSP\footnote{Which are the correct inputs for the base of elements in stellar population fitting using the {\sc starlight} code.}, where the contribution of each stellar spectral type is weighted by the number and luminosity of the stars. This is a large task in its own right and is well beyond the scope of this paper. This means that although we can reproduce the mean spectrum with both approaches, the results need to be taken with caution. AGB and regular giant stars are short-lived, and their true contribution to a stellar population spectrum of a given age cannot be determined by the approaches we described here.

\begin{figure}
%\epsscale{1.2}
\includegraphics[scale=0.55]{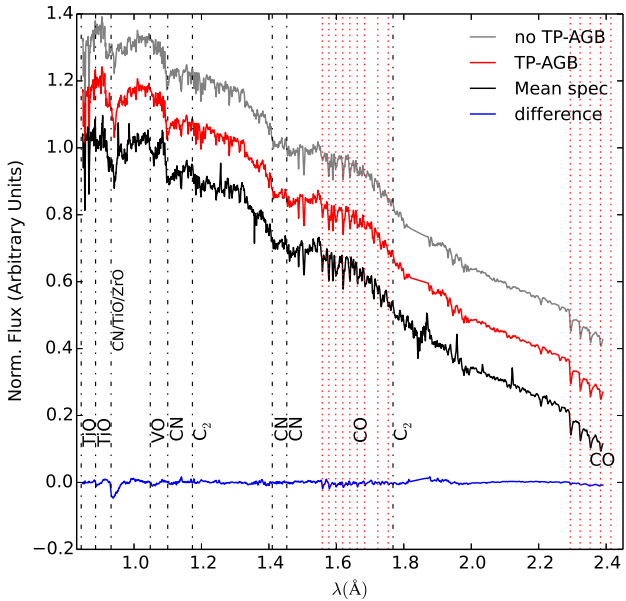}
\caption{Difference between the {\it stars} and {\it stars no-TP-AGB} approaches presented in Fig.~\ref{hr}.}
\label{difs}
\end{figure}

With that said, TP-AGB stars are very luminous. Despite the fact that they exist only in population components with ages $\sim$0.2 - 2 Gyr (M05), they have been shown to account for almost 20\% of the 1.6 $\mu$m flux of nearby galaxies  \citep{melbourne12}, with much of the remainder likely coming from relatively featureless main sequence stars. In the K-band they may be responsible for roughly half the total luminosity \citep{salaris14}.  Such luminous stars, with strong absorption features, may be expected to leave an imprint on the NIR spectrum. Also, there is a subtle difference between the two fits in the absorption band at 0.93 $\mu$m (see Fig.~\ref{difs}). As noted above, this band probably contains contributions from CN, TiO, and ZrO. The fit in this region is improved when TP-AGB stars are included, suggesting that some of the absorption indeed comes from the Zr brought to the surface by the third dredge-up. Combined with the fact that VO is present in the galaxies' spectra (see Fig.~\ref{meanConts}, these considerations suggest that TP-AGB stars play a significant role in the features observed in our integrated spectra\footnote{We note that the VO feature near 1.05 $\mu$m is not well reproduced by either the models or the stellar library. The absence of the feature in the model spectra is not surprising, as they do not include the VO molecule. Its weakness in the empirical fit may be a result of incompleteness of the stellar library or the fact we are not constraining it by the rules of stellar evolution. A definite conclusion will require more detailed analysis, and is left for a future paper (R. Riffel et al {in preparation})}.

 If TP-AGB stars dominate the NIR spectral features, we might expect to observe stronger features in intermediate-age galaxies that should host larger populations of these stars. This is not obviously the case in this sample. The spectra in Figure \ref{sources} are remarkably similar, and there is no clear relationship between the EW values in Table \ref{ew} and whether a galaxy is classified as intermediate-age or old in Table \ref{props}. Of the galaxies in this sample, NGC~5005 has probably the most secure evidence for a large intermediate-age ($\sim$45\%) stellar population \citep{cid04b}. Its CN/TiO/ZrO and VO bands, however, are the weakest that we observe (Table \ref{ew}). Similarly,  \citet{Mason15} examined three galaxies with intermediate-age populations and found them to have weaker molecular absorption bands than older galaxies.

There are several possible reasons for this apparent contradiction. First, stellar populations derived from optical and NIR spectroscopy cannot necessarily be directly compared. Dust can obscure young populations in the optical, and different slit sizes\footnote{That might be the case of NGC5005, since the slit used by \citet{cid04b} is three times wider than ours, thus potentially sampling additional stellar populations.} can probe different regions of spatially varied populations \citep{riffel_passp,martins13}. Second, the intermediate-age galaxies studied here and in \citet{Mason15} likely span only limited range of metallicity. As the strength of TP-AGB spectral features depends greatly on their age and metallicity (M05), weak NIR features like those observed in NGC~5005 may not be a universal characteristic of intermediate-age galaxies. Third, intermediate-age galaxies may also contain young stellar populations, whose relatively featureless spectra may dilute the stronger bands of the TP-AGB stars.

Finally, as shown by our model fits, red giants, C-R and E-AGB stars can also produce most of the observed features sometimes attributed to TP-AGB stars. These stars may be important in stellar populations that are both younger than the intermediate-age populations in which TP-AGB stars are believed to be abundant (if R C-rich and E-AGB stars are present in large fractions), or in older populations (which contain large numbers of red giants). This clearly demonstrates that to accurately model a stellar population, all kinds of evolved stars need to be taken into account.

%We search for correlations between the EW and other galaxies properties like H$_{\alpha}$ luminosity (wich can be related with the star formation rate) and the optical ages for the stellar populations. No clear correlations were found. In the first case, it may be due to the fact that the stars emitting ionizing photons are very hot (younger) and the absorption bands are due to colder (older) stars. Regarding the latter, it is probability due to extinction and spatially-varying stellar populations, reinforcing the fact that stellar population ages derived from optical and near-IR spectroscopy are not necessarily directly comparable \citep{riffel_passp,martins13}. One way to better address this point is to study the stellar populations of the same galaxies, from the optical to near-infrared using the same apertures. Since these kind of data are not available yet, such analysis is behind the scope of this paper.

To date, the most targeted search for TP-AGB stars in galaxies is that of \citet{zibetti13}. They obtained NIR (H and K-band) spectra of 16 post-starburst galaxies, selected partly for the CN (1.41 $\mu$m) and C$_2$ (1.77 $\mu$m) bands to be redshifted out of strong telluric absorption (z$\sim$0.2). Although the stellar ages (0.5 - 1.5 Gyr) of these galaxies correspond to the period of maximum contribution of TP-AGB stars to the integrated spectrum, the depth of absorption features from carbon molecules in these stars was found to be no more than a few per cent of the continuum.  Extrapolating from our observations to predict the features expected in post-starburst galaxies is beyond the scope of this paper, but we note that the weak CN and C$_2$ absorptions reported here would not be detected at the resolution ($\sim$300 {\it versus} 1200) and S/N ($\sim$30 vs 110 in the K band) of the \citet{zibetti13} data, and their spectral coverage does not include most of the other features that we report. It is therefore difficult to gauge whether our interpretation of the molecular features as dominated by TP-AGB stars is in conflict with the observations of \citet{zibetti13}. However, it is clear that the features in the post-starburst galaxies are not as strong as predicted by the M05 models for galaxies of those ages and metallicities.

\subsection{Performance of the models}

The last two approaches tested in \S\ref{ssp} (M11 approach and BC03 approach) are intended to show how well different SSP recipes are able to reproduce these features. 
Fitting the mean spectrum with the M11 models results in a complex star formation history, with similar fractions of young (t $<$ 1\,Gyr), intermediate-age (t $\sim$ 1\,Gyr) and old (t $\sim$ 10\,Gyr) SSPs. With the BC03 models, though, {\sc starlight} includes a young ($\sim10^{7}$ yr) component and requires only a very small fraction of old SSPs. This is probably because the BC03 models do not have deep NIR CN bands, causing the code to attempt to fit the 1.1 \mc\ CN band with Pa$\gamma$. If such a young component were genuinely present, we would expect to require a significant fraction of F stars (the hottest available in the IRTF library) when fitting the mean spectrum with the stellar library, which is not the case. Furthermore, if we remove the $\leq$100~Myr components in the BC03 approach, the best fit is achieved with $\sim$ 35\% of the light due to the 1~Gyr and $\sim$ 65\% due to the 2~Gyr components, but with a much poorer fit ($\chi^2 \sim$~100). 

The high ($\sim$65\%) light fraction of stars with ages 0.2 -- 3 $\times$10$^9$ yr in the M11 model fits ostensibly supports our previous interpretation that the molecular bands likely contain a strong contribution from TP-AGB stars, which should dominate the NIR emission of a stellar population of this age. 
However, this result could be coincidental: the M11 models include empirical spectra with good spectral resolution only for the TP-AGB stars, meaning only the intermediate age stellar population in their models will present these features. As discussed above, most of the features can be reproduced with a mix of E-AGB and R C-rich stars, which means older stellar populations probably also have these features at some level. To avoid misinterpreting NIR galaxy spectra, it is thus of great importance to include high quality empirical spectra of the other evolutionary phases in the EPS models.

While the M11 models qualitatively reproduce several of the broad absorption bands in the mean spectrum (notably CO and CN), these bands are weaker or absent in the BC03 model fit. This is probably not related to the different spectral resolution of the models ($R \sim 200 $, vs $R \sim 500$ in M11) because we find that similar results are obtained from the M11 and M05 models (the latter with $R \sim 250 $ in the K-band). The differences are more likely related to uncertainties in the theoretical treatment of AGB and TP-AGB stellar phases and the fact that the M05/M11 models include empirical spectra of C and O-rich stars and therefore better account for these features.

Since the relative band strengths depend on metallicity (M05), we also tried fitting the mean spectrum with BC03 and M05 models including three non-solar metallicities (the M11 models include only solar
metallicity for this wavelength range). Similar ages to those derived from the original analysis were obtained in all cases, and changing the metallicity did not significantly improve the fits to the features. This may suggest that the incompleteness in stellar types/phases of the empirical spectra on which the models are based is the first order problem in reproducing the NIR absorption features. At this stage in the development of the models, metallicity probably plays a secondary role in predicting the strength of the features

\section{Final Remarks}\label{rem}

We have presented high-S/N, 0.85 -- 2.5 $\mu$m spectra of the nuclei of 12 local galaxies. The spectra show indications of numerous weak molecular absorption bands that arise in the atmospheres of cool stars. While several types of evolved stars can exhibit most of the detected features, we argue that the high luminosity of TP-AGB stars, and the fact that the spectra contain VO and likely ZrO, that are only expected in stars undergoing thermal pulses, imply that these features are the long-debated signatures of TP-AGB stars. However, a combination of R-type carbon rich stars and E-AGB stars should not be ruled out as an important contributor to these bands.

In testing two sets of stellar population synthesis models against these spectra, we found that models that include stars with relatively strong absorption bands provide a more accurate representation of the spectra. However, to avoid misinterpreting the data, future models must include high-quality spectra of C-R and E-AGB stars, along with those of TP-AGB stars.
Testing the predictions of EPS models has to date been hampered by a lack of high-quality data againstwhich the models can be tested, and we hope that the high S/N and wavelength coverage of the spectra presented here will make them a useful aid to the development of such models. The reduced spectra presented here are available at: http://www.canfar.phys.uvic.ca/vosui/\#/karun/xdgnirs\_Dec2014/XD\_final.

\section*{Acknowledgements}
 We thank the anonymous referee for useful comments. The Brazilian authors thank CNPq and FAPERGS support. RR acknowledges M. Cushing, L. Girardi,  C. Maraston, A. Romero and R. P. Schiavon
for helpful discussions as well as CNPq and FAPERGS for financial support. LCH acknowledges support by the Chinese Academy of Science through grant No. XDB09030102 (Emergence of Cosmological Structures) from the Strategic Priority Research Program and by the National Natural Science Foundation of China through grant No. 11473002. CRA is supported by a Marie Curie Intra European Fellowship within the 7th European Community Framework Programme (PIEF-GA-2012-327934). Based on observations obtained
at the Gemini Observatory, operated by the Association of
Universities for Research in Astronomy, Inc., under a cooperative
agreement with the NSF on behalf of the Gemini partnership: the
National Science Foundation (US), the Science and
Technology Facilities Council (UK), the National Research
Council (Canada), CONICYT (Chile), the Australian Research Council
(Australia), Minist\'{e}rio da Ci\^{e}ncia e Tecnologia (Brazil)
and Ministerio de Ciencia, Tecnolog\'{i}a e Innovaci\'{o}n Productiva (Argentina).

\end{document}